\documentclass{iopjournal}

\usepackage{amsmath}
\usepackage{bm}
\usepackage{mathtools}
\usepackage{amssymb}
\newcommand{\blue}[1]{{#1}}

\begin{document}

\articletype{Focus on Insights from kHz Gravitational Waves} 

\title{Quarkyonic matter and hadron-quark crossover from an ultracold atom perspective}

\author{Hiroyuki Tajima$^{1,2,3,*}$\orcid{0000-0001-5247-7116}, Kei Iida$^{4,2}$\orcid{0000-0002-9088-1109},
Toru Kojo$^5$\orcid{0000-0001-5656-3652},
and Haozhao Liang$^{1,3,6}$\orcid{0000-0002-2950-8559}}

\affil{$^1$Department of Physics, Graduate School of Science, The University of Tokyo, Tokyo 113-0033, Japan}

\affil{$^2$RIKEN Nishina Center, Wako 351-0198, Japan}

\affil{$^3$ Quark Nuclear Science Institute, The University of Tokyo, Tokyo 113-0033, Japan}

\affil{$^4$ Department of Liberal Arts, The Open University of Japan, Chiba 261-8586, Japan}

\affil{$^5$ Theory Center, IPNS, High Energy Accelerator Research Organization (KEK), 1-1 Oho, Tsukuba, Ibaraki 305-0801, Japan}

\affil{$^6$ RIKEN Center for Interdisciplinary Theoretical and Mathematical Sciences, Wako 351-0198, Japan}

\affil{$^*$Author to whom any correspondence should be addressed.}

\email{hiroyuki.tajima@tnp.phys.s.u-tokyo.ac.jp}


\begin{abstract}
The dense matter equation of state is of great interest due to the recent development of astrophysical observations for neutron stars.
A rapid increase in pressure indicates a continuous crossover from a hadron phase to a quark phase without any phase transitions, yet its microscopic mechanism remains elusive. Recently, a peak in the speed of sound and a baryon momentum-shell structure, which are predicted from a quarkyonic matter picture, have been regarded as key features of the hadron-quark crossover.
In this work, we explore a field-theoretical framework to describe the hadron-quark crossover, drawing an analogy with the Bose-Einstein condensate to Bardeen-Cooper-Schrieffer (BEC-BCS) crossover established in ultracold atomic experiments. 
Strikingly, a peak in the speed of sound and the baryon momentum-shell structure can simultaneously be explained by the tripling fluctuation effect arising from a different context of quantum many-body physics.
We demonstrate these properties in a simplified model and provide a microscopic derivation of the quarkyonic matter model within our field-theoretical framework.
\end{abstract}

\section{Introduction}
The fate of extremely dense matter has been a long-standing and important pending problem.
Thanks to recent progress in astrophysical observations of neutron stars, this problem is now directly relevant to neutron star physics, although the lattice simulation of quantum chromodynamics (QCD) suffers from a severe sign problem for dense matter~\cite{fukushima2010phase}.
In particular, the observation of the mass-radius relation enables us to extract the equation of state (EOS) in dense matter that exists inside a massive neutron star~\cite{ozel2016masses,li2021progress}.

So far, it has still been debated how baryon matter changes to quark matter with increasing density in the core region of a neutron star.
While various scenarios, including the first-order phase transition, have been proposed~\cite{fukushima2010phase,Baym_2018}, a promising candidate, called the hadron-quark crossover~\cite{Masuda:2012kf},
where baryon matter changes to quark matter without any phase transitions, has attracted attention based on recent astrophysical observations.
The hadron-quark crossover scenario originates from the hadron-quark continuity~\cite{PhysRevLett.82.3956}, and the crossover EOS is constructed by an interpolation between the nuclear matter EOS in the low-density regime and the weakly-coupled quark matter EOS~\cite{Masuda:2012kf}.
An important feature of the hadron-quark crossover is a peak in the density dependence of the speed of sound, which originates from the disparity of the EOS between baryon matter and quark matter~\cite{kojo2021qcd}.
 This characteristic behavior is consistent with recent astrophysical observations of the mass-radius relation~\cite{PhysRevD.110.034035}.
Moreover, the hadron-quark crossover in the binary neutron star merger will be detected in the future kilohertz gravitational-wave observations~\cite{PhysRevLett.129.181101,PhysRevLett.130.091404}.

However, the microscopic origin of the peaked speed of sound has been elusive.
In this context, a quarkyonic matter picture~\cite{McLerran:2007qj}, in which baryon and quark degrees of freedom coexist in the high-density regime with the large-$N_{\rm c}$ limit, gives a possible explanation of the hadron-quark crossover. 
In Ref.~\cite{PhysRevLett.122.122701}, the hadron-quark crossover EOS was formulated as a quarkyonic matter EOS with residual baryonic correlations near the quark Fermi surface.
Although physical parameters are introduced phenomenologically, the resulting EOS describes the peaked speed of sound in the crossover regime. 
This quarkyonic matter model has been 
sophisticated microscopically in Ref.~\cite{PhysRevLett.132.112701} by exploiting an explicit quark-hadron duality.
Although direct evidence for these states is still lacking,
it is a promising research direction to explore the microscopic nature of the quarkyonic hadron-quark crossover from different points of view.

To this end, it is useful to consider another crossover phenomenon established experimentally.
In an ultracold Fermi gas, 
by tuning the attractive interaction via the Fano-Feshbach resonance~\cite{RevModPhys.82.1225},
one can realize a crossover from
the Bose-Einstein condensate (BEC) of tightly bound molecules to
the weakly-interacting Bardeen-Cooper-Schriffer (BCS) Fermi superfluidity~\cite{zwerger2011bcs,strinati2018bcs,OHASHI2020103739}.
The BEC-BCS crossover has also been realized in superconducting materials by tuning the carrier density~\cite{kasahara2014field,nakagawa2021gate,PhysRevX.12.011016} \blue{(see also theoretical studies of the density-induced BEC-BCS crossover~\cite{PhysRevB.60.12410,shi2022density,PhysRevA.106.043308,sakakibara2023finite}).
The neutron-proton pairing in nuclear matter can also be regarded as another example of the density-induced BEC-BCS crossover~\cite{PhysRevC.79.034304,PhysRevC.81.034007,PhysRevC.82.024911,PhysRevC.90.065804,tajima2019superfluid}. Moreover, the}
recent lattice simulation of dense two-color quantum chromodynamics (QC$_2$D) indicates the density-induced BEC-BCS crossover, exhibiting the peaked speed of sound~\cite{iida2022velocity,iida2024two}.
Accordingly, the diquark BEC-BCS crossover and EOS in dense QC$_2$D matter have been investigated within the mean-field theory~\cite{PhysRevD.105.076001,PhysRevD.109.076006,PhysRevD.111.094006}.

A crucial difference between two- and three-color systems is the quantum statistics of baryons.
While the mean-field approach can describe the BEC-BCS crossover associated with bosonic baryons,
it is difficult to describe the formation of fermionic baryons and their density-induced crossover in the mean-field approach.
In such a case, fluctuation effects play a crucial role as in the case of the BEC-BCS crossover above the critical temperature where the mean-field order parameter vanishes~\cite{nozieres1985bose}.
To overcome this difficulty,
in Ref.~\cite{PRL135.042701_4ywp-752m},
tripling fluctuations, which are analogous to pairing fluctuations in the BEC-BCS crossover, have been taken into account to see the microscopic physics of the crossover concerning the fermionic baryon formation, within the phase-shift representation of clustering fluctuations~\cite{Dashen:1969ep}.
The spectral properties of tripling fluctuations across the density evolution have been studied in terms of the Green's function technique~\cite{PhysRevResearch.4.L012021,tajima2023density}.

In this paper, we review the tripling fluctuation theory for the hadron-quark crossover developed in Ref.~\cite{PRL135.042701_4ywp-752m}.
In particular, we show that the characteristic features of quarkyonic matter model in the crossover regime,
that is, the baryonic momentum shell
and the peak speed of sound~\cite{PhysRevLett.122.122701,PhysRevLett.132.112701}, are explained by the field-theoretical approach developed from a different context.
Moreover, we show the microscopic derivation of the quarkyonic matter model developed in Ref.~\cite{PhysRevLett.122.122701} in terms of the tripling fluctuation theory.

This paper is organized as follows.
In Sec.~\ref{sec:2}, we show the field-theoretical formalism for $N$-body clustering fluctuations on thermodynamic quantities, where $N=3$ corresponds to the case of tripling fluctuations.
In Sec.~\ref{sec:3}, we apply the tripling fluctuation theory to a simplified non-relativistic 1D model, which enables us to track fluctuation physics analytically.
The application to 3D relativistic matter is also discussed.
Finally, we summarize this paper in Sec.~\ref{sec:4}.
Throughout the paper, we take $\hbar=k_{\rm B}=1$ and the system's volume is regarded as unity.

\section{Field-theoretical approach to tripling  fluctuations}
\label{sec:2}
We start from the phase-shift representation of the $N$-body clustering fluctuations induced by the short-range $N$-body interaction.
The grand-canonical thermodynamic potential can be written as
\begin{align}
\label{eq:1}
    \Omega=\Omega_{0}+\sum_{N=2}^{\infty}\delta\Omega_{N},
\end{align}
where $\Omega_{0}$ is the non-interacting term and
\begin{align}
\label{eq:2}
    \delta\Omega_N=-T\sum_{\bm{K}}\sum_{n}
    \ln\left[1-V_NG_0(\bm{K},i\omega_n)\right]
\end{align}
is the correction term associated with $N$-body clustering fluctuations~\cite{Dashen:1969ep}.
In Eq.~\eqref{eq:2},  
the interaction
$V_{N}$ is responsible for the formation of an $N$-body cluster and is assumed to be a short-range (i.e., its momentum dependence is negligible).
$G_0(\bm{K},i\omega_n)$ is the bare $N$-body propagator with the total momentum $\bm{K}$ and Matsubara frequency $\omega_n$, where all relative momenta and Matsubara frequencies are summed up. 
For $N=2$, Eq.~\eqref{eq:2} recovers the pairing fluctuation term for the BEC-BCS crossover, developed by Nozi\`{e}res and Schmitt-Rink~\cite{nozieres1985bose}(which is equivalent to Gaussian fluctuations in terms of the path integral formalism~\cite{PhysRevLett.71.3202}).
For fermionic $N$-body clusters, the Matsubara frequency is fermionic as $\omega_n=(2n+1)\pi T$ ($n\in \mathbb{Z}$),
and the summation of $\omega_n$ can be converted to the real-frequency integration as~\cite{PRL135.042701_4ywp-752m}
\begin{align}
\label{eq:3}
   \delta\Omega_N=-\sum_{\bm{K}} 
   \int_{-\infty}^{\infty}\frac{d\omega}{\pi}
   f(\omega-N\mu)\phi(\bm{K},\omega),
\end{align}
where $f(\omega)=(e^{\omega/T}+1)^{-1}$ is the Fermi distribution function,
$\mu$ is the chemical potential of a constituent particle,
and $\phi(\bm{K},\omega)$
is the phase shift of the $N$-body propagator.
A crucial difference between the present case of fermionic clusters and that of bosonic clusters~\cite{nozieres1985bose} is whether the Fermi and Bose distribution functions appear in the expression of $\delta\Omega_N$. 
The dressed retarded $N$-body propagator $G(\bm{K},\omega)$ is related to the bare one $G_0(\bm{K},i\omega_n\rightarrow\omega+i\eta)$ (where $\eta$ is an infinitesimally positive small number) as
\begin{align}
\label{eq:4}
    \frac{G(\bm{K},\omega)}{G_0(\bm{K},\omega)}=e^{i\phi(\bm{K},\omega)}\left|
    \frac{G(\bm{K},\omega)}{G_0(\bm{K},\omega)}
    \right|.
\end{align}
The interaction effect associated with $N$-body cluster formation is incorporated into $\phi(\bm{K},\omega)$.
In this way, one can address thermodynamic quantities such as fermion number density $\rho=-\frac{\partial\Omega}{\partial\mu}$ across the density-induced crossover.
In particular, the case of $N=3$ corresponds to tripling fluctuations~\cite{PRL135.042701_4ywp-752m}.
We note that while the Hartree-Fock term is subtracted in the fluctuation term of Ref.~\cite{PRL135.042701_4ywp-752m},
this does not make significant differences for the short-range interactions.

Generally, $\phi(\bm{K},\omega)$ satisfies 
\begin{align}
\label{eq:const}
    \phi(\bm{K},\omega\rightarrow\pm \infty)=0,
\end{align}
because of the sum rule
\begin{align}
\label{eq:6-2}
    \int_{-\infty}^{\infty}d\omega\,{\rm Im}G(\bm{K},\omega)=\int_{-\infty}^{\infty}d\omega\,{\rm Im}G_0(\bm{K},\omega).
\end{align}
If the system involves an $N$-body bound state
$\phi(\bm{K},\omega)$ exhibits a positive $\pi$ shift at a negative energy.
On the other hand, Eq.~\eqref{eq:const} indicates that $\phi(\bm{K},\omega)$ should be reduced back to zero and thus $\phi(\bm{K},\omega)$ decreases at some interval of $\omega$ beyond the continuum threshold.
The interplay of the bound state at a negative energy and the scattering state at positive energy leads to the non-monotonic momentum distribution of $N$-body fermionic clusters, as we shall demonstrate in a toy model.

\section{Demonstration of tripling fluctuation effects in a simplified model}
\label{sec:3}
The formulation of tripling fluctuation effects presented in Sec.~\ref{sec:2} can be used in both non-relativistic and relativistic systems with arbitrary spatial dimensions.
In this section, we elucidate the role of tripling fluctuations in a non-relativistic model for the sake of its qualitative understanding.
Moreover, we provide the microscopic derivation of the quarkyonic  matter model~\cite{PhysRevLett.122.122701} empirically connecting non-relativistic baryon matter and relativistic quark matter.

\subsection{1D non-relativistic model}
Here we present how tripling fluctuations appear in the density-induced hadron-quark crossover through the demonstration in a non-relativistic 1D model exhibiting asymptotic freedom and trace anomaly~\cite{PhysRevLett.120.243002,daza2019quantum,PhysRevA.100.063604,PhysRevA.102.023313}.
We consider three-color fermions with a short-range attractive three-body interaction described by a Lagrangian density
\begin{align}
\label{eq:5}
    \mathcal{L}= \sum_{\alpha={\rm r,g,b}}\bar{\psi}_\alpha\left(\partial_\tau-\frac{\partial_x^2}{2m}-\mu\right)\psi_\alpha -V_3
    \bar{\Psi}\Psi,
\end{align}
where $\psi_\alpha$ and $\bar{\psi}_\alpha$ are fermionic Grassmann fields with a mass $m$, a chemical potential $\mu$, and a color index $\alpha={\rm r,g, b}$.
In the second term of Eq.~\eqref{eq:5},
$V_3$ is the short-range three-body coupling strength, and $\bar{\Psi}=\bar{\psi}_{\rm r}\bar{\psi}_{\rm g}\bar{\psi}_{\rm b}$ and
$\Psi=\psi_{\rm b}\psi_{\rm g}\psi_{\rm r}$
represent the baryon fields.

Based on the ladder-type resummation,
we obtain
\begin{align}
\label{eq:6}
    G(K,\omega)=\frac{G_0(K,\omega)}{1-V_3G_0(K,\omega)},
\end{align}
where
\begin{align}
    G_0(K,\omega)
    =\sum_{k,q}\frac{Q(k,q,K)
    }{\omega+i\eta-\varepsilon_{q/2+K/3+k}-\varepsilon_{K/3-q}-\varepsilon_{q/2+K/3-k}}
\end{align}
is the bare three-body propagator with
the Pauli-blocking factor $Q(k,q,K)$~\cite{PhysRevResearch.4.L012021,PhysRevC.109.055203} and
the kinetic energy $\varepsilon_k=k^2/2m$.
For simplicity, we employ the in-vacuum counterpart (i.e., $Q(k,q,Q)\rightarrow 1$ for the intermediate states) to obtain the analytical expression~\cite{PRL135.042701_4ywp-752m,PhysRevLett.120.243002}
\begin{align}
\label{eq:8}
    G_0(K,\omega)\simeq -\frac{m}{2\sqrt{3}\pi}\ln\left(\frac{\tilde{\omega}_K+i\eta-\Lambda^2/m}{\tilde{\omega}_K+i\eta}\right),
\end{align}
where
$\tilde{\omega}_K=\omega-K^2/2M_{\rm B}$ (where $M_{\rm B}=3m$), and
$\Lambda$ is the ultraviolet momentum cutoff.
Using Eqs.~\eqref{eq:6} and \eqref{eq:8} at $\mu=0$,
we obtain the three-body binding energy $\mathcal{B}$ \blue{from the pole condition~\cite{PhysRevLett.120.243002,hryhorchak2025bipolaron}
\begin{align}
   0&= 1-V_3G_0(K=0,\omega=-\mathcal{B})\cr
   &=1+\frac{mV_3}{2\sqrt{3}}\ln\left(\frac{\Lambda^2}{m\mathcal{B}}\right),
\end{align}
where we take $\Lambda\rightarrow \infty$.
In this way, we obtain}
\begin{align}
    \mathcal{B}=\frac{\Lambda^2}{m}e^{\frac{2\sqrt{3}\pi}{mV_3}}.
\end{align}
In this regard, $V_3$ can be expressed in terms of $\mathcal{B}$ and \blue{thus $\Lambda$ dependence of physical quantities is eliminated by this renormalization procedure}.

Using Eqs.~\eqref{eq:4} and \eqref{eq:6}, we obtain the phase shift
\begin{align}
    \phi(K,\omega)=\pi\theta(\tilde{\omega}_K+\mathcal{B})\theta(-\tilde{\omega}_K)
    +\theta(\tilde{\omega}_K)\tan^{-1}\left[\frac{\pi}{\ln(|\tilde{\omega}_K|/\mathcal{B})}\right],
\end{align}
where $\theta(x)$ is the step function.
Within the tripling fluctuation theory, the fermion number density $\rho$ is given by
\begin{align}
    \rho&=-\frac{\partial \Omega_0}{\partial\mu}-\frac{\partial\delta\Omega_3}{\partial\mu}\cr
    &\equiv \rho_{\rm Q} +\rho_{\rm B},
\end{align}
where 
\begin{align}
    \rho_{\rm Q}=3\sum_{k}f_Q(k)\equiv 3\sum_{k}f(\varepsilon_k-\mu)
\end{align}
is a non-interacting fermion density (corresponding to the quark number density in the sense of the analog quantum simulation for dense matter), and
\begin{align}
    \rho_{\rm B}=3\sum_{K}f_{\rm B}(K)
\end{align}
is the tripling fluctuation term that gives a baryon-like trimer density in the dilute limit.
Note that $\rho$ corresponds to the total number density of quark-like constituent fermions, and thus the net baryon density is related to $\rho/3$.
The baryonic momentum distribution $f_{\rm B}(K)$ is given by
\begin{align}
    f_{\rm B}(K)=\int_{-\infty}^{\infty}d\omega \,f(\omega-3\mu)\left[A(K,{\omega})-A_0(K,\omega)\right],
\end{align}
where
\begin{align}
\label{eq:17}
A(K,{\omega})-A_0(K,\omega)&=\frac{1}{\pi}\frac{\partial}{\partial\tilde{\omega}_K}\phi(K,\omega)\cr
    &\equiv \delta(\tilde{\omega}_K+\mathcal{B})-\frac{\theta(\tilde{\omega}_K)}{\tilde{\omega}_K\left[\ln(\tilde{\omega}_K/\mathcal{B})+\pi^2\right]},
\end{align}
is the difference between the correlated three-particle excitation spectrum $A(K,\omega)$ and the free-streaming one $A_0(k,\omega)=-\frac{1}{\pi}{\rm Im}\,G_0(K,\omega)$, and $\delta(x)$ is the delta function.
Importantly, Eq.~\eqref{eq:17} consists of positive and negative contributions associated with the bound state and the scattering state.
More explicitly, we obtain
\begin{align}
\label{eq:18}
    f_{\rm B}(K)=f\left({K^2}/{2M_{\rm B}}-3\mu-\mathcal{B}\right)
    -\int_0^{\infty}\frac{d\omega}{\omega}\frac{f\left({K^2}/{2M_{\rm B}}-3\mu+\omega\right)}{\ln(\omega/\mathcal{B})+\pi^2},
\end{align}
where the first and second terms correspond to the bound-state and scattering-state contributions, respectively.

\begin{figure}[t]
    \centering
    \includegraphics[width=0.85\linewidth]{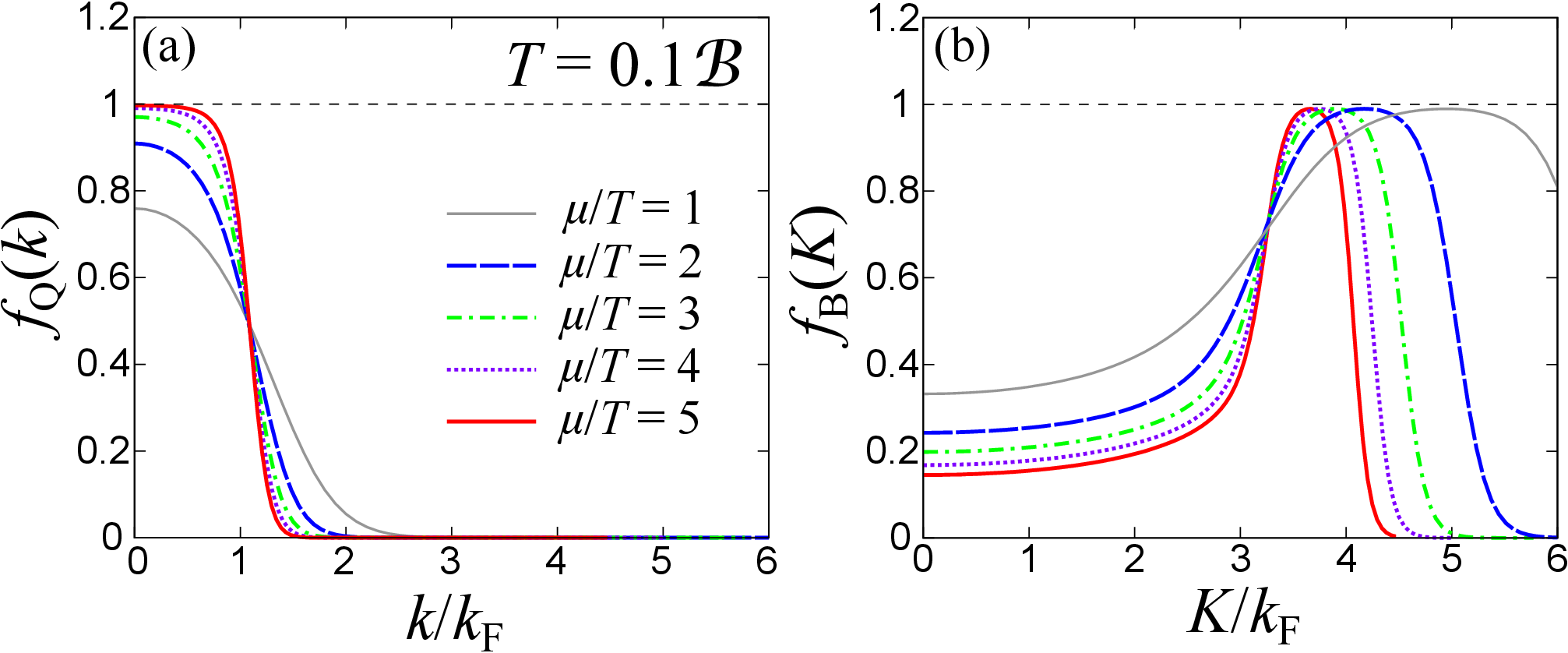}
    \caption{Calculated momentum distributions of (a) quark-like fermions $f_{\rm Q}(k)$ and (b) baryon-like trimers $f_{\rm B}(K)$
    at several $\mu/T$,
    where $k_{\rm F}=\sqrt{2m\mu}$ is the Fermi momentum. The temperature is fixed at $T=0.1\mathcal{B}$. }
    \label{fig:1}
\end{figure}

Figure~\ref{fig:1} shows the numerical results of $f_{\rm Q}(k)$ and $f_{\rm B}(K)$ at $T=0.1\mathcal{B}$.
While $f_{\rm Q}(k)$ monotonically increases with increasing $\mu$ and gradually approaches the Fermi-step behavior at larger $\mu$,
$f_{\rm B}(k)$ shows the momentum shell structure $f_{\rm B}(K\neq 0)\simeq 1$ around $K\gtrsim 3k_{\rm F}$ with the Fermi momentum $k_{\rm F}=\sqrt{2m\mu}$,
which is reminiscent of the baryonic momentum shell in the quarkyonic matter picture~\cite{PhysRevLett.130.091404}.
At small $K$, the cancellation between the bound and scattering contributions in Eq.~\eqref{eq:18} leads to the strong suppression of $f_{\rm B}(K)$.
On the other hand, just above $\blue{K}=3k_{\rm F}$, the negative scattering-state contribution disappears, and thus only the positive bound-state contribution survives at the baryonic momentum shell region.
Eventually, $f_{\rm B}(K)$ becomes zero in the limit of $K\rightarrow\infty$.
At $T\rightarrow 0$, the width $\Delta_{\rm B}$ of the baryonic momentum shell can be obtained as~\cite{PRL135.042701_4ywp-752m}
\begin{align}
    \Delta_{\rm B}&=\sqrt{(3k_{\rm F})^2+2M_{\rm B}\mathcal{B}}-3k_{\rm F}\cr
    &=\frac{M_{\rm B}\mathcal{B}}{3k_{\rm F}}+O(M_{\rm B}^2\mathcal{B}^2/k_{\rm F}^3).
\end{align}
Moreover, the finite-temperature effect is taken into account in our approach.
One can find that the baryonic momentum shell is smeared at smaller $\mu/T$.
\blue{While we neglect the Pauli-blocking effect in the evaluation of the tripling fluctuation term, such an effect may reduce the in-medium three-body binding energy~\cite{PhysRevResearch.4.L012021} and hence $\Delta_{\rm B}$. However, the qualitative behavior would be unchanged by the modification of the binding energy.}

\blue{We note that the cancellation between bound and scattering state contributions may appear in the pairing fluctuation term (see., e.g., Ref.~\cite{PhysRevA.77.023626}) in dense QC$_2$D matter. Meanwhile, it should be noted that the quantum statistics of clusters make significant differences in their momentum distributions. While three-color clusters (i.e., trimer) exhibits the Fermi-step behavior, two-color clusters (i.e., dimer) obey the Bose statistics where the low-momentum distribution is strongly enhanced at low temperatures. In this sense, the cancellation effect might be masked by the Bose enhancement. In addition, the mean-field bosonic condensate plays a crucial role in the QC$_2$D case~\cite{PhysRevD.105.076001}.}

\begin{figure}[t]
    \centering
    \includegraphics[width=0.5\linewidth]{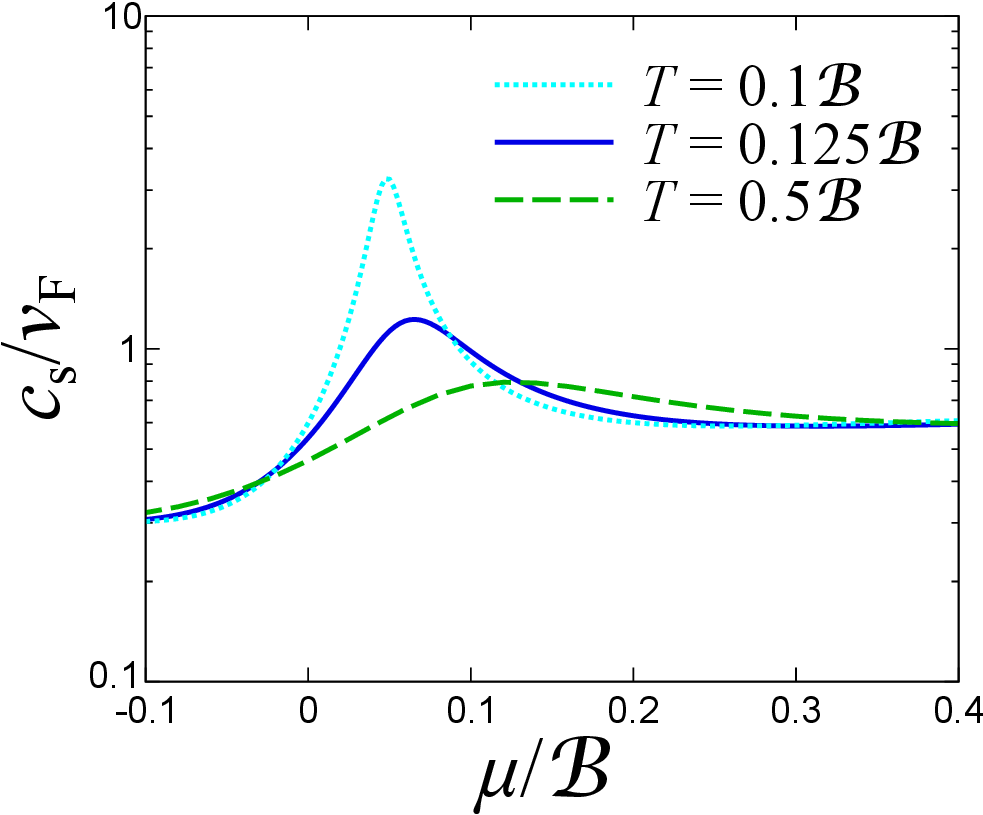}
    \caption{Isothermal speed of sound $c_{\rm s}$ normalized by the Fermi velocity $v_{\rm F}=k_{\rm F}/m$. \blue{The temperatures are taken as $T=0.1\mathcal{B}$, $0.125\mathcal{B}$ and $0.5\mathcal{B}$.}}
    \label{fig:2}
\end{figure}

Figure~\ref{fig:2} shows the isothermal speed of sound $c_{\rm s}$, which is given by
\begin{align}
    c_{\rm s}=\sqrt{\frac{\rho}{m}\left(\frac{\partial \rho}{\partial\mu}\right)_T^{-1}},
\end{align}
in a non-relativistic system.
Remarkably, $c_{\rm s}$ exhibits a peak structure in the evolution of $\mu$.
This behavior can be understood as the suppression of the density susceptibility $\chi=\left(\frac{\partial\rho}{\partial\mu}\right)_T$, which can be rewritten as
\begin{align}
\label{eq:21}
    \chi=3\sum_{k}\frac{\partial}{\partial\mu}f_{\rm Q}(k)
    +3\sum_{K}\frac{\partial}{\partial\mu}f_{\rm B}(K).
\end{align}
While the first term in Eq.~\eqref{eq:21} is always positive as the density susceptibility of free fermions,
the second term in Eq.~\eqref{eq:21} can be negative as we find that the small momentum part of $f_{\rm B}(K)$ is suppressed with increasing $\mu$ in Fig.~\ref{fig:1}.
Therefore, $\chi$ becomes small because two terms in Eq.~\eqref{eq:21} cancel with each other in the crossover regime, leading to the enhancement of $c_{\rm s}$ that is proportional to $\chi^{-1/2}$. 
At high density limit, $c_{\rm s}$ approaches the Fermi velocity $v_{\rm F}=k_{\rm F}/m$ corresponding to the non-relativistic conformal limit.
\blue{At higher temperatures, the peak of $c_{\rm s}$ is gradually smeared out (see also Ref.~\cite{PRL135.042701_4ywp-752m}). At nonzero temperatures, $c_s$ is nonzero even at $\mu\leq 0$ due to the thermal pressure in the non-relativistic system. In the dilute limit (i.e., $\mu/T\rightarrow-\infty$), our result reproduces the third virial expansion exactly~\cite{PRL135.042701_4ywp-752m,PhysRevLett.120.243002}.}

We note that our result overestimates tripling fluctuation effects on thermodynamic quantities compared to the quantum Monte Carlo simulation~\cite{PhysRevA.102.023313}.
Our calculation could be more accurate if the in-medium effect on $G_0(K,\omega)$ is considered.
However, note that the Mott effect~\cite{ropke1983particle,Blaschke:2013zaa}, where the bound state is suppressed by the Pauli-blocking effect at low $K$, is absent in the present 1D system~\cite{PhysRevResearch.4.L012021}.
Although the in-medium shift of $\mathcal{B}$ might exist, this would not affect the qualitative behavior of $f_{\rm B}(K)$. 
Moreover, the peak behavior of $c_{\rm s}$ (in other words, the minimum of $\chi$) can be found in Ref.~\cite{PhysRevA.102.023313}.

\blue{Before ending this section, we mention the current status of cold atomic studies toward quantum simulation of three-color fermions. 
While it is known that strongly-interacting three-component Fermi gases are unstable due to strong three-body losses~\cite{OHara_2011}, a recent experimental study reveals an anomalous loss process in the three-dimensional system~\cite{schumacher2026observation}, motivating us to study the underlying microscopic physics specific to three-component fermions. Moreover, our toy model originally proposed in Ref.~\cite{PhysRevLett.120.243002} involves the three-body interaction, which is one of the challenging topics in ultracold atoms. 
So far, there are several theoretical proposals~\cite{PhysRevA.69.043607,PhysRevA.90.021601,PhysRevA.97.011602,PhysRevA.109.013319,PhysRevA.111.053319,v1y3-2b6r}
and indeed the tunable three-body interaction has been realized experimentally~\cite{PhysRevLett.128.083401}. Based on these circumstances, the experimental demonstration of the three-body crossover might be realized in future experiments.}

\subsection{Microscopic description of 3D relativistic quarkyonic matter}
Here we discuss the connection to the relativistic quarkyonic matter model~\cite{PhysRevLett.122.122701}.
The phase-shift approach to $N$-body clustering is valid even for relativistic systems~\cite{Dashen:1969ep} and the antiparticle contribution can be negligible at sufficiently high densities.
For simplicity, we ignore the baryon-baryon and quark-quark interactions except for the confinement force~\cite{PhysRevLett.132.112701}.
While quarks behave as relativistic particles, baryons are assumed to be non-relativistic particles.
In such a case, we obtain the net baryon number density $\rho/N_{\rm c}$ in symmetric matter (where $N_{\rm c}=3$ is the color degrees of freedom) as
\begin{align}
\label{eq:22}
    \frac{\rho}{N_{\rm c}}=4\sum_{\bm{k}}f\left(\sqrt{k^2+m_{\rm Q}^2}-\mu\right)+4\sum_{\bm{K}}\int_{-\infty}^{\infty}
    d\omega\, 
    \left[A_{\rm B}(\bm{K},\omega)-A_0(\bm{K},\omega)\right]f\left(\omega-N_{\rm c}\mu\right),
\end{align}
where $m_{\rm Q}$ is the constituent quark mass, $A_{\rm B}(\bm{K},\omega)$ is the fluctuating baryon spectrum (which represents repeated dissociation and formation of baryons in medium), and $A_0(\bm{K},\omega)$ is the non-interacting three-quark excitation spectrum.
Note that the factor $4$ in Eq.~\eqref{eq:22} originates from the isospin degree of freedom.
Although $A_{\rm B}(\bm{K},\omega)$ should be obtained from microscopic QCD physics~\cite{RevModPhys.82.1095},
it is reasonable to consider $A_{\rm B}(\bm{K},\omega)$ consisting of the baryonic pole and the correction on the $N_{\rm c}$-quark scattering continuum as
\begin{align}
    A_{\rm B}(\bm{K},\omega)
    -A_{0}(\bm{K},\omega)
    &\simeq\delta\left(\omega-\frac{K^2}{2M_{\rm B}}-N_{\rm c}m_{\rm Q}+\mathcal{B}\right) \cr
    &\quad-\theta\left(\omega-\frac{K^2}{2M_{\rm B}}-N_{\rm c}m_{\rm Q}\right)\delta A_{\rm scatt}(\bm{K},\omega),
\end{align}
where $\mathcal{B}$ is the binding energy due to the strong interaction and
$\delta A_{\rm scatt}(\bm{K},\omega)>0$ is the scattering correction on the continuum.
At low densities where $m_{\rm Q}-\mathcal{B}/N_{\rm c}<\mu<m_{\rm Q}$, Eq.~\eqref{eq:22} yields
\begin{align}
    \frac{\rho}{N_{\rm c}}\simeq 4\sum_{\bm{K}}f\left(\frac{K^2}{2M_{\rm B}}-\mu_{\rm B}\right),
\end{align}
with $\mu_{\rm B}=N_{\rm c}(\mu-m_{\rm Q})+\mathcal{B}>0$, indicating that the system is dominated by baryons.
On the other hand, at $\mu\geq m_{\rm Q}$ and $T/\mu\simeq 0$,
the quark distribution and the scattering continuum contribution become simultaneously nonzero as
\begin{align}
\label{eq:25}
    \frac{\rho}{N_{\rm c}}&\simeq \frac{2k_{\rm FQ}^3}{3\pi^2}
    +\frac{2k_{\rm FB}^3}{3\pi^2}
    -4\sum_{\bm{K}}\int_{0}^{N_{\rm c}\mu-N_{\rm c}m_{\rm Q}-\frac{K^2}{2M_{\rm B}}}d\omega \,\delta A_{\rm scatt}\left(\bm{K},\omega+\frac{K^2}{2M_{\rm B}}+N_{\rm c}m_{\rm Q}\right),
\end{align}
where $k_{\rm FQ}=\sqrt{\mu^2-m_{\rm Q}^2}$ and $k_{\rm FB}=\sqrt{2M_{\rm B}\mu_{\rm B}}\equiv\sqrt{2M_{\rm B}(N_{\rm c}\mu-N_{\rm c}m_{\rm Q}+\mathcal{B})}$ are the quark and baryon Fermi momenta, respectively.
Although we do not know the exact form of $\delta A_{\rm scatt}(\bm{K},\omega)$, it is obvious that the scattering-state contribution is active only for $|\bm{K}|\leq \sqrt{2M_{\rm B}N_{\rm c}(\mu-m_{\rm Q})}$ based on the interval of the $\omega$ integration.
For simplicity, suppose that the momentum integrand of the third term in 
Eq.~\eqref{eq:25} is given by
\begin{align}
\label{eq:26}
    \int_{0}^{N_{\rm c}\mu-N_{\rm c}m_{\rm Q}-\frac{K^2}{2M_{\rm B}}}d\omega \,\delta A_{\rm scatt}\left(\bm{K},\omega+\frac{K^2}{2M_{\rm B}}+N_{\rm c}m_{\rm Q}\right)
    \simeq \theta\left(\sqrt{2M_{\rm B}N_{\rm c}(\mu-m_{\rm Q})}-|\bm{K}|\right),
\end{align}
which is deduced from the assumption that the spectral frequency sum rule given by Eq.~\eqref{eq:6-2} is exhausted at $\omega\leq N_{\rm c}\mu$ as
\begin{align}
    \int_{-\infty}^{N_{\rm c}\mu}d\omega\,[A_{\rm B}(\bm{K},\omega)-A_0(\bm{K},\omega)]\simeq 0,
\end{align}
for $|\bm{K}|<\sqrt{2M_{\rm B}N_c(\mu-m_Q)}$.
Using Eqs.~\eqref{eq:25} and \eqref{eq:26}, we obtain
\begin{align}
\label{eq:27}
    \frac{\rho}{N_{\rm c}}\simeq 
    \frac{2k_{\rm FQ}^3}{3\pi^2}
    +\frac{2k_{\rm FB}^3}{3\pi^2} -\frac{2(k_{\rm FB}-\Delta_{\rm B})^3}{3\pi^2},
\end{align}
where
\begin{align}
    \Delta_{\rm B}=\sqrt{2M_{\rm B}(N_{\rm c}\mu-N_{\rm c}m_{\rm Q}+\mathcal{B})}-\sqrt{2M_{\rm B}N_{\rm c}(\mu-m_{\rm Q})}
\end{align}
corresponds to the width of the baryon momentum shell.
Eq.~\eqref{eq:27} is nothing more than the net baryon number density in Ref.~\cite{PhysRevLett.122.122701}.
In this way, we establish the microscopic foundation of the quarkyonic matter model from an ultracold atom perspective, i.e., the tripling fluctuation theory.
Note that the approximation given by Eq.~\eqref{eq:26} does not necessarily hold in a realistic situation, as we demonstrated in a toy model (where the residual contribution survives below $K=3k_{\rm F}$).
This contribution is also found in the IdylliQ (ideal dual quarkyonic) matter model~\cite{PhysRevLett.130.091404}.
Nevertheless, the essential feature of the quarkyonic EOS is sufficiently described even under this simplification as shown in Ref.~\cite{PhysRevLett.122.122701}.

\blue{Our result gives an insight for improving the quarkyonic matter models based on other bases, such as the extended relativistic mean-field model~\cite{xia2018nuclear,PhysRevD.108.054013} and the extended Nambu-Jona-Lasinio model~\cite{PhysRevD.110.014022,cao2025extended}. While most of the previous works consider only the bound-state pole contribution for baryons, our result suggests that the scattering state contribution plays a crucial role in reproducing the baryonic momentum shell structure microscopically.}

\section{Summary}
\label{sec:4}

We have explored the microscopic mechanism of the quarkyonic matter and the hadron-quark crossover from a viewpoint of ultracold atomic physics, where the BEC-BCS crossover has been well established both theoretically and experimentally.
In particular, we have investigated the role of fluctuating baryonic correlations (repeated process of dissociation and formation) by exploiting the phase-shift representation of $N$-body clustering fluctuations, in analogy with the Nozi\`{e}res--Schmitt-Rink approach for the BEC-BCS crossover.  

Through the analysis in the simplified model, we have elucidated that the tripling fluctuation theory explains the essential features of the quarkyonic matter and the hadron-quark crossover, that is, the baryonic momentum shell and the peaked speed of sound. The interplay between the bound- and scattering-state contributions in baryon-like tripling fluctuations leads to the strong suppression of the low-momentum baryon distributions.
The enhancement of the speed of sound in the intermediate regime can be understood as a consequence of the non-monotonic evolution of the momentum distributions.
The relationship between the present framework and the quarkyonic matter model has also been clarified.

While we perform the numerical demonstration in the simplified model, it is worth applying our approach to realistic neutron-star matter.
Since our approach includes the finite-temperature effect, it would be useful for the future investigation of high-energy astrophysical phenomena such as binary neutron-star mergers and core-collapsed supernovae.
It is also interesting to compare our scenarios with models that exhibit exotic ordered phases~\cite{PhysRevD.71.114006,PhysRevA.104.L041302,95dn-656y}.

\ack{The authors thank Joaqu\'{i}n E. Drut, Yaqi Hou, and Eiji Nakano for useful discussion.}

\funding{This work was supported in 
part by Grants-in-Aid for Scientific Research provided by 
JSPS through Grants No.~JP18H05406, No.~JP22H01158, No.~JP22K13981, No.~JP23H01167, No.~JP23K25864, and No.~JP25K01001.}





\bibliographystyle{iopart-num}

\bibliography{reference}

\end{document}